\documentclass[twocolumn,11pt]{article}
\usepackage{epsfig}
\usepackage{amsfonts}
\usepackage{amssymb}
\usepackage{amsmath}
\usepackage{bm}

\usepackage{mathptm}
\usepackage{mathrsfs}

\usepackage{color}

\newcommand{\R}{{\mathbb R}}

\newcommand\ZJ[1]{\mathchoice
                 {{\buildrel{\hspace*{.1em}{_{\,\boldsymbol\circ}}}\over{#1}}}
                 {{\buildrel{\hspace*{.1em}{_{\,\boldsymbol\circ}}}\over{#1}}}
                 {{\buildrel{\hspace*{.1em}{\boldsymbol\circ}}\over{#1}}}
                 {{\buildrel{\hspace*{.1em}{\boldsymbol\circ}}\over{#1}}}}

\newcommand\DT[1]{\mathchoice
  {{\buildrel{\hspace*{.1em}\text{\Large\bf.}}\over{#1}}}
    {{\buildrel{\hspace*{.1em}\text{\Large\bf.}}\over{#1}}}
                 {{\buildrel{\hspace*{.1em}\text{\Large.}}\over{#1}}}
                 {{\buildrel{\hspace*{.1em}\text{\large.}}\over{#1}}}}
\newcommand\Ee{{\bm E}}              
\newcommand\Ep{{\bm\varPi}}              
\newcommand\EE{{\bm e}}
\newcommand{\lineunder}[2]{\LU{\begin{array}[t]{c}\underbrace{#1}\vspace*{.5em}\end{array}}{\mbox{\footnotesize\rm #2}}}
\newcommand{\LU}[2]{\begin{array}[t]{c}#1\vspace*{-1em}\\_{#2}\end{array}}
\newcommand{\linesunder}[3]{\LSU{\begin{array}[t]{c}\underbrace{#1}\vspace*{.5em}\end{array}}{\mbox{\footnotesize\rm #2}}{\mbox{\footnotesize\rm #3}}}
\newcommand{\LSU}[3]{\begin{array}[t]{c}#1\vspace*{-1em}\\_{#2}\vspace*{-.5em}\\_{#3}\end{array}}
\renewcommand{\d}{{\rm d}}
\usepackage{psfrag}
\usepackage{epsfig}
\newcounter{myfigure}
\newenvironment{my-picture}[3]{\refstepcounter{myfigure}\label{#3}\setlength{\unitlength}{\textwidth}\begin{picture}(#1,#2)}{\end{picture}}
\usepackage{cite}
\newcommand\pdt[1]{\frac{\partial{#1}}{\partial t}}
\newcommand\Frakg{\text{\large$\mathfrak{g}$}}

\begin{document}
\begin{sloppypar}

\title{\LARGE{\bf
Thermodynamically consistent model for poroelastic rocks\hfill\hspace*{1em}\\
towards tectonic and volcanic processes and earthquakes
\hfill\hspace*{1em}}}
\date{}
\author{\Large\hspace*{-.5em} T. Roub\'{\i}\v cek\hfill\hspace*{1em}\medskip\baselineskip10pt
\\\small\hspace*{-.3em}Institute of Thermomechanics, Czech Academy of Sciences, Dolej\v skova 5, CZ-182 00 Praha 8, \hfill\hspace*{1em}
\\[-.1em]\small\hspace*{-.3em}and\hfill\hspace*{1em}
\\[-.1em]\small\hspace*{-.3em}Mathematical Institute, Charles University, Sokolovsk\'a 83, 
CZ-186~75~Praha~8, Czech Republic\hfill\hspace*{1em}
\\\\\\
\hspace*{-.5em}\begin{minipage}[t]{1\textwidth}\baselineskip10pt
  {\small {\bf Summary.}
 A general-purpose model combining concepts from rational continuum
    mechanics, fracture and damage mechanics, plasticity, and poromechanics is
    devised in Eulerian coordinates, involving objective time derivatives.
    The model complies with
    mass, momentum, and energy conservation as well as entropy inequality
    and objectivity.
    It is devised to cover many diverse phenomena, specifically
rupture of existing lithospheric faults, tectonic earthquakes, generation 
and propagation of seismic waves, birth of new tectonic faults,
or volcanic activity, aseismic creep, folding of rocks, aging of rocks,
long-distance saturated water transport and flow in poroelastic rocks, 
melting of rocks and formation of magma chambers, or solidification of magma.
}
\\\\\\
\end{minipage}
}

\maketitle

\baselineskip14pt





\def\vv{{\bm v}}
\def\uu{{\bm u}}
\def\xx{{\bm x}}
\def\yy{{\bm y}}
\def\nn{{\bm n}}
\def\ff{{\bm g}}
\def\zz{\chi}
\def\EE{{\bm e}}
\def\bbC{{\mathbb C}}
\def\bbD{{\mathbb D}}
\def\bbI{{\mathbb I}}
\def\bbK{K}
\def\bbM{M}
\def\HEAT{h}
\newcommand{\bmE}{\bm{\varTheta}}
\newcommand{\GM}{G_\text{\sc m}^{}}
\def\W{\vartheta}
\def\w{w}
\def\DIF{D}

\section{Introduction -- phenomena and\\processes to be modelled
}\label{sec1}


Earth's upper mantle and crust exhibit multi-physics
multi-scale character in various senses whose modelling is extremely
challenging and was given an intensive attention
for decades. Purely elastodynamical or visco-elastodynamical
mechanical processes are combined with various inelastic
processes, with thermodynamical effects including  solid-fluid 
phase 
 transition, and with transport processes. This will
be the focus of a multi-purpose model  
devised in this paper, while other phenomena as
chemistry or magnetism will be ignored. 

On long time scales, the phenomena of aseismic creep,
folding of rocks, long-distance water saturated flow and
transport in poroelastic rocks, aging of rocks,
melting of rocks and formation of magma chambers, or
solidification of magma are to be considered.
On short time scales, the  phenomena to be considered
are rupture of existing lithospheric
faults, tectonic earthquakes, generation 
of seismic waves and propagation through nonhomogeneous
solid and fluidic areas, birth of new tectonic faults,
or volcanic activity including volcanic earthquakes.

Usually, only models focused to particular phenomena
are cast in geophysics, which is partly dictated
by computational implementability. Nevertheless,
in reality, the above mentioned phenomena are mutually
coupled and influence each other. Therefore,
building a model which would capture many (or all)
of these processes is of a theoretical interest.

Although the elastic strains remain always rather
small as materials (rocks) cannot withstand too much
large elastic strain without triggering inelastic
processes (damage or plastification),
the displacement can be surely large within
millions of years and even solid rocks behave
fluidic on long time scales. An important aspect
is that there is no reference configuration,
in contrast to engineering workpieces where a
reference configuration may refer to the shape
in which they have been manufactured and the
Lagrangean description is well motivated even for
largely strained elastic solids. In contrast,
the planet Earth is in a constant manufacturing
process. This suggests to employ rather the Eulerian
description and small-strain models  in
a rate form  combined with a properly devised transport.
 The rate formulation is related with the concept of
hypoelasticity.  Of course, thermodynamical consistency as
the mass, the momentum, and the energy conservations as well as the 
entropy inequality are ultimate attributes
to be  respected  by each rational model.
 Involving a mere elastic stress rate would however
lead to corruption of energy conservation in a closed deformation
cycle, as realized already by \cite{TruNol65NFTM}.
This gives rise to an extra (here called ``structural'') stress
which will balance the energetics of the model, cf.\ ${\bm S}_{\rm str}$
in \eqref{eq2} below.

The basic ingredients to be involved are proper rheological visco-elastic
models and proper choice of internal variables. Particular modelling
techniques from fracture (or damage) mechanics, plasticity, poromechanics,
and thermomechanics are to be naturally combined. It seems that no model of
such a physical consistency and a generality has been devised so far in
geophysical literature, although several particular models
of this sort have been devised and largely used, even though sometimes lacking
full mechanical or thermodynamical consistency, cf.\ also 
Section~\ref{sec-concl}.  A very inspiring attempt
has been done in \cite{LyHaBZ11NLVE}, however without using objective
derivatives for transport of tensorial variables and with incomplete
structural stress and continuity equation avoided, as well as with mixing
state-dependent free energy with rate-dependent dissipation potential
(so that the energy balance and also entropy inequality cannot be satisfied).
Besides, \cite{LyHaBZ11NLVE} is focused to tectonic applications, while here
also a wider context (like volcanic) will be considered.
Let us still remark that an attempt to formulate the model \cite{LyHaBZ11NLVE}
thermodynamically consistently has been done
in \cite{Roub17GMHF} in terms of displacements,
but not all time-derivatives have been objective
and the energy balance was thus corrupted.
An improved variant using the velocity-strain
formulation and convective times derivatives
was devised in \cite{RouTom??CMPE},
although only isothermal and not fully objective.
Several aspects of the model have been devised
in \cite{Roub??SPTC} towards solid-liquid phase
transformation as occurs in freezing of water
and melting of ice in a simplified semi-compressible
variant (like in Sect.\,\ref{sec-semi}), which can be
here exploited for solidification of magma and melting
of rocks.

Some 
damageless variants have been devised in  
 \cite{Dint13SCST,DiPrGe19SZUM,PopSob08TTDT} or,
 small-strain non-convective isothermal,
 in \cite{RoSoVo13MRLF,RouVod19MMPF}.
 Actually, there is an ample literature on various multi-phase,
multi-physics models for geodynamics
\cite{BeRiSc01TPMC,KelSuc19CMMP,YarPod15DPVM,Gery19INGM}
and references therein, including fully coupled
seismo-hydro-mechanical models (e.g., \cite{PGYD20SHMM}
and references therein) and models for magma dynamics
(e.g., \cite{KeMaKa13MMMD} and references therein); cf.\ 
also Section~\ref{sec-concl}. 
 
 Also fully large-strain models have been
devised, too. In Eulerian formulation
the deformation-gradient, or distortion, or the
left Cauchy-Green tensors are transported,
cf.\ \cite{GCTP21UFOH,GodPes10TCNM,Rubi19EFIM,TCRG20STAA,Volo13AELD}.
As an alternative  possibility, one  should also 
mention 
a reference-coordinate (Lagrangian)
formulation at large strains in \cite{RouSte18TEPR},
using an artificial reference configuration and quite
heavy push-forward and pull-back 
machinery as well as a hidden spurious hardening in large
slips, as articulated later in \cite{DaRoSt??NHFV}.  Such
geometrically fully nonlinear model using the Kr\"oner-Lee
multiplicative decomposition of the deformation gradient
may still benefit from the small elastic strain concept,
as relevant in particular in rock mechanics. 

The goal of this article is to devise a thermodynamically
consistent model which will be fully convective (i.e.\
in actual Eulerian coordinates) employing
objective time derivatives and which will 
cover all the above mentioned phenomena.
We will use (in contrast to \cite{LyHaBZ11NLVE})
the conventional concept of free energy which
does not include rates together with 
a dissipation potential of nonconservative forces.

The plan is to formulate a general model in
Sections~\ref{sec-nom}--\ref{sec-model}, together with some
specification towards modelling poroelastic rock and magma,
including solid-fluid phase transition in Section~\ref{sec-specific}.
Usage of the model for earthquake and some volcanic processes modeling
is discussed in Sections~\ref{sec-EQ} and \ref{sec-PT}, respectively.
Then, in Section~\ref{sec-termodynam}, the general model is
scrutinized as far as its thermodynamics concerns. Eventually,
Section~\ref{sec-semi} briefly mentions some simplification for
media which are only slightly compressible  and
Section~\ref{sec-concl} gives some summary  and some comparison 
with other existing approaches and some outlook for computer
implementations. 





\section{Nomenclature}\label{sec-nom}
Let us first summarize the main notation concerning variables
and thermodynamical potentials of our model:
  
\vspace*{-.2em}\hspace*{2em}$\varrho\ \ $ mass density,

\vspace*{-.2em}\hspace*{2em}$\vv\ \ $ velocity,

 \vspace*{-.2em}\hspace*{2em}$\w=(\alpha,\phi)\ \ $ with $\alpha$ damage and $\phi$ porosity,

 \vspace*{-.2em}\hspace*{2em}$\Ee\ \ $ elastic strain (symmetric),

   \vspace*{-.2em}\hspace*{2em}$\Ep\ \ $ inelastic strain (symmetric deviatoric),
  
   \vspace*{-.2em}\hspace*{2em}$\zz\ \ $ water (or oil) content,

  \vspace*{-.2em}\hspace*{2em}$\theta\ \ $ temperature,

 \vspace*{-.2em}\hspace*{2em}$\W=\gamma(\theta)\ \ $ enthalpy,

 \vspace*{-.2em}\hspace*{2em}$\bm{S}_\text{\sc e}^{}\ \ $ the elastic stress tensor (symmetric),

\vspace*{-.2em}\hspace*{2em}$\bm{S}_\text{\sc v}^{}\ \ $ the viscous stress tensor (symmetric),

\vspace*{-.2em}\hspace*{2em}$\bm{S}_{\rm str}^{}\ \ $ a structural stress
(symmetric),

  \vspace*{-.2em}\hspace*{2em}$\eta\ \ $ entropy,

  \vspace*{-.2em}\hspace*{2em}$\psi\ \ $ free energy,

 \newpage

 \vspace*{-.2em}\hspace*{2em}$\mu\ \ $ a chemical potential (here pore pressure),

 \vspace*{-.2em}\hspace*{2em}$\zeta\ \ $ dissipation potential,
  
  \vspace*{-.2em}\hspace*{2em}$\xi\ \ $ heat production rate,
  
\vspace*{-.2em}\hspace*{2em}$\EE(\vv)=
  {\rm sym}(\nabla\vv)\ \ $ total-strain rate,

\vspace*{-.2em}\hspace*{2em}$\bm j\ \ $ the heat flux ($\,=-K\nabla\theta$, i.e.\ Fourier's law).
  


\noindent
Further notation concerns basic material parameters and outer
conditions or sources:

 \vspace*{-.2em}\hspace*{2em}$K_\text{\sc e}^{}$, $G_\text{\sc e}^{}$
  elastic bulk and shear modulus (in Pa),
  
 \vspace*{-.2em}\hspace*{2em}$K_\text{\sc v}^{}$, $G_\text{\sc v}^{}$
viscous bulk and shear modulus (in Pa\,s),


\vspace*{-.2em}\hspace*{2em}$G_\text{\sc m}^{}\ \ $ Maxwellian (creep) shear modulus (in Pa\,s),

\vspace*{-.2em}\hspace*{2em}$G_\text{\sc y}^{}\ \ $ plastification yield stress (in Pa),

\vspace*{-.2em}\hspace*{2em}$\DIF\ \ $ viscosity modulus for diffusion (in Pa\,s),

\vspace*{-.2em}\hspace*{2em}$\bbM\ \ $ mobility coefficient of diffusant (water of oil),

\vspace*{-.2em}\hspace*{2em}$B$, $\beta\ \ $ Biot modulus and coefficient,

\vspace*{-.2em}\hspace*{2em}$\bbK\ \ $ heat-conductivity coefficient,

\vspace*{-.2em}\hspace*{2em}$\ff\ \ $ bulk (gravitational) acceleration,

\vspace*{-.2em}\hspace*{2em}$r$\ \ bulk (radiogenic) heat source,


\vspace*{-.2em}\hspace*{2em}$\bm t\ \ $ traction force,

  \vspace*{-.2em}\hspace*{2em}$j_{\rm ext}^{}\ \ $ an external heat flux.
  
  \noindent
For a 3$\times$3-matrices $A$, we will use the 
decomposition $A={\rm sym}\,A+{\rm skew}\,A$ with
${\rm sym}\,A=\frac12A^\top\!+\frac12A$ and ${\rm skew}\,A
=\frac12A-\frac12A^\top$ and also  the orthogonal decomposition 
$A={\rm sph}\,A+{\rm dev}\,A$ with the spherical (volumetric) part
${\rm sph}\,A=({\rm tr}\,A)\bbI/3$ and the deviatoric part 
${\rm dev}\,A=A-({\rm tr}\,A)\bbI/3$, where 
$\bbI$ denotes the identity matrix.

\section{A general-purpose model}\label{sec-model}
The departure point is the {\it additive}, also called
{\it Green-Naghdi's, decomposition} of the total small strain
\begin{align}\label{GND}
\EE(\uu)=\Ee+\Ep
\end{align}
into the elastic and the inelastic parts, where $\uu$ denotes the displacement.
Consistently with the rate formulation of the model, also  
this decomposition will be expressed in
rates, cf.\ \eqref{eq3} below, so that the displacement will be 
eliminated from the model. The inelastic strain $\Ep$ is a
deviatoric (i.e.\ trace-free) tensor in a position of an internal variable,
which captures creep or plastic-like effects
which are typically isochoric (i.e.\
does not affect volumetric purely viscoelastic variations).

We will use the conventional notation $(\cdot)'$ for (partial)
derivatives of functions. Further, the {\it convective} (material)
{\it time derivative} and the {\it Zaremba-Jaumann}
{\it corotational derivative}  \cite{Jaum11GSPC,Zare03FPTR}
will be denoted, respectively, by:
\begin{subequations}\begin{align}\label{eq:19}
&\DT{(\cdot)}=\Big[\pdt{}+{\bm v}\cdot\nabla\Big](\cdot)\ \ \ \text{ and }\ \ \
    \\&\ZJ{(\cdot)}=\DT{(\cdot)}+(\cdot){\bm W}-{\bm W}(\cdot)
    \ \ \ \text{ with }\ \ \
  {\bm W}={\rm skew}\,\nabla\vv\,,
\label{ZJ}\end{align}\end{subequations}
where ${\bm W}$ is the {\it spin tensor} of motion. The objective derivative
\eqref{ZJ} is used for tensors, and in particular it is justified for stress
rates by 
\cite[p.494]{Biot65MID}, cf.\ also \cite{Fial11GSSM}. For
isotropic materials, also the strain rates as exploited in our model below,
can be justified \cite[Rem.\,1]{Roub??SPTC}. 
This corotational objective time derivative is also  (perhaps most) often
 used in geophysical modelling,
viz \cite{BabSob08HRNM,BeuPod10VMCL,DiPrGe19SZUM,DDGP18SBMB,GerYue07RCMM,HeGeDi18IRSD,JacCac20MMBD,KeMaKa13MMMD,MQLM07CASN,PCTC17SMFV,PGYD20SHMM,PopSob08TTDT,ThKaPo15LSRB,YarPod15DPVM}
or \cite[Chap.12]{Gery19INGM} and references therein. It allows for a proper
modelling of situations when the medium rotates within rock folding/bending or
within magma vortices.
Another justification refers to the example \eqref{stored} below:
if the shear elastic modulus $G_\text{\sc e}^{}$ vanishes, the stress tensor
degenerate to a pressure and the viscoelastic solid material becomes a
viscoelastic fluid which, if further the bulk modulus $K_\text{\sc e}^{}$ goes to
$\infty$, becomes the standard Navier-Stokes viscous incompressible
fluid, cf. \cite[Rem.\,2]{Roub??SPTC}. This would not hold for other
objective (like Oldroyd's or Green-Naghdi's or Lie's) derivatives.
The important consequence of this corotation time derivative choice
is that
\begin{align}\label{ZJ-property}
{\rm tr}\ZJ\Ep=({\rm tr}\Ep)\!\DT{^{}}\quad\text{ and }\quad
\ZJ\Ep^\top=(\Ep^\top)\!\ZJ{^{}}
\end{align}
so the flow-rule \eqref{eq4} below keeps $\Ep$ as symmetric trace-free within
time evolution provided the initial condition for $\Ep$ has these attributes,
and then \eqref{eq3} keeps $\Ee$ symmetric. 

The basic ingredients of the model are the {\it free energy}
considered here, for simplicity, in a decoupled form as
the sum of the stored energy $\varphi$ and the heat part:
\begin{align}\label{free}
  \psi(\Ee,\w,\zz,\nabla\w,\theta)=
  \varphi(\Ee,\w,\zz)+\frac\kappa2|\nabla\w|^2+\HEAT(\theta)
\end{align}
and a {\it dissipation potential} (or pseudopotential
of dissipative forces), assumed here additively (de)coupled as
\begin{align}\nonumber
  &\zeta(\w,\theta;\EE,\ZJ\Ep,\DT\w,\nabla\mu,\DT\zz,\nabla\ZJ\Ep)=
\zeta_1(\EE)+\zeta_2(\w,\theta;\ZJ\Ep)
  \\&\qquad\ 
+\zeta_3(\w,\theta;\DT\w)+
\frac12\bbM|\nabla\mu|^2
+\frac12\DIF\DT\zz^2
+\frac\varkappa2|\nabla\ZJ\Ep|^2\,.
\label{dissip-pot}\end{align}
The corresponding dissipation rate (=\,heat production rate) is then
\begin{align}\nonumber
&\xi(\w,\theta;\EE,\ZJ\Ep,\DT\w,\nabla\mu,\DT\zz,\nabla\ZJ\Ep)=
\zeta_1'\big(\EE\big){:}\EE
\\&\nonumber\qquad\quad
+\big[\zeta_2\big]_{\ZJ\Ep}'\big(\w,\theta;\ZJ\Ep\big){:}\ZJ\Ep
+[\zeta_3]_{\DT\w}'\big(\w,\theta;\DT\w\big)\cdot\DT\w
\\&\qquad\qquad\qquad\qquad
+\bbM|\nabla\mu|^2+\DIF\DT\zz^2+\varkappa|\nabla\ZJ\Ep|^2.
\label{dissip-rate}
\end{align}

We devise the system of seven equations, namely: mass transport (continuity)
equation \eqref{eq1}, momentum equation \eqref{eq2}, Green-Naghdi's
decomposition in rates \eqref{eq3}, flow rule for inelastic strain \eqref{eq4},
flow rule for internal variables (damage and porosity) \eqref{eq5},
diffusion of diffusants (typically water or, in petrology, oil) \eqref{eq6},
and heat transfer equation \eqref{eq7}.
Specifically,
\begin{subequations}\label{eq}\begin{align}\label{eq1}
&\DT\varrho=-\varrho{\rm div}\,\vv\,,
\\\nonumber
&\varrho\DT\vv={\rm div}\big(\bm{S}_\text{\sc e}^{}
         +{\bm S}_{\rm v}
+{\bm S}_{\rm str}\big)
+
\varrho\ff\,
  \\\nonumber
  &\ \text{ with }{\bm S}_\text{\sc e}^{}=\varphi_\Ee'(\Ee,\w,\zz)\,,\ \ \
      {\bm S}_\text{\sc v}^{}=
\zeta_1'(\EE(\vv))\,,\ \text{ and }
 \\
 &\ \ {\bm S}_{\rm str}=
 \kappa\nabla\w{\otimes}\nabla\w
  -\Big(\varphi(\Ee,\w,\zz){+}\frac{\kappa}2|\nabla\w|^2\!{+}\HEAT(\theta)\Big)\bbI\,,\!\!\!\!
\label{eq2}\\[-.3em]\label{eq3}
&   \ZJ{\Ee}+\ZJ{\Ep}=\EE(\vv)\,,
    \\
    &
    \big[\zeta_2\big]_{\ZJ\Ep}'(\w,\theta;\ZJ\Ep)+{\rm dev}\bm{S}_\text{\sc e}^{}
=\varkappa\Delta\ZJ{\Ep}\,,
  \label{eq4}  \\
  &
  [\zeta_3]_{\DT w}'\big(\w,\theta;
    \DT{\w}\big)+\varphi_\w'(\Ee,\w,\zz)=\kappa\Delta\w\,,
   \label{eq5} \\
&\DT\zz={\rm div}(\bbM\nabla\mu)
   \ \ \ \text{ with }\ \ \ \mu=\DIF\DT\zz+
   \varphi_\zz'(\Ee,\w,\zz)\,,
    \label{eq6}\\\nonumber
 &\DT\W={\rm div}\big(\bbK\nabla\theta\big)
    +\xi(\w,\theta;\EE(\vv),\ZJ\Ep,\DT\w,\nabla\mu,\nabla\ZJ\Ep)
\\&\hspace{3.3em}
+
(\HEAT(\theta){-}\W){\rm div}\,\vv+r
\ \ \ \ \text{ with }\ \ \ \ \W=\gamma(\theta)
\,,
\label{eq7}\end{align}\end{subequations}
where  $\gamma$ is related to $\HEAT$  from  \eqref{free} by 
\begin{align}
\gamma 
(\theta)=\HEAT(\theta)-\theta\HEAT'(\theta).
\end{align}
In particular,
${\gamma\hspace{.2em}}'(\theta)=-\theta\HEAT''(\theta)$ is the
temperature-dependent {\it heat capacity}.
The specific form of the
structural stress ${\bm S}_{\rm str}$ in \eqref{eq2}
is important to balance the energy and will be derived in detail in
Sect.\,\ref{sec-termodynam} below.
The additive decomposition \eqref{GND} is now written in
rates as \eqref{eq3}. It is used in geophysical models in a non-objective
variant e.g.\ in \cite{BilHir07RCSD,ReLYue03MSZG}
and in the objective corrotation derivative variant in
\cite{DDGP18SBMB,DiPrGe19SZUM,DDGP18SBMB,Gery19INGM,GerYue07RCMM}.

Actually, in bigger spatial domains of interest and
longer time scales, the radiogenic heat $r$ in \eqref{eq7}
varies considerably in space and time.
A more detailed modelling of a spatially inhomogeneous
and decaying-in-time heat source would still
need transport equations for relevant radioisotopes
contributing to $r$, specifically
\begin{align}
  r=\sum_ir_i
\ \ \ \ \text{ with }\ \ \ 
  \tau_i\DT r_i+r_i=0\,,
\end{align}
where $\tau_i$ denotes the half-life-time constants for particular
radioisotopes (specifically for $^{232}$Th, $^{238}$U, and $^{40}$K).

The coefficient $\varkappa$ in \eqref{eq4} determines
a dynamical length scale of possible slip zones (cataclastic
zones in tectonic faults). It is important that
this gradient term involves the rate $\ZJ\Ep$ but not
$\Ep$ itself; actually, involving
gradient of $\Ep$ would cause a spurious hardening effects
during long lasting slips, as articulated in \cite{DaRoSt??NHFV}.
The coefficient $\kappa$ in \eqref{eq5}
determines length scales of particular scalar
internal variables. In particular, the gradient of damage
variable $\alpha$ influences a typical width of the
damage zone around the core of tectonic faults, as devised in
\cite{LyHaBZ11NLVE}. Actually, $\kappa$ can be 
a vector of coefficients applying on particular components
of $\w$; for notational simplicity, we write it
here just as a single coefficient.

Roughly speaking, the philosophy of the 
mechanical part of the model
(\ref{eq}b--f) with $\varrho$ constant and with \eqref{eq6} written in the
form $\Delta_{\bbM}^{-1}\DT\zz-\DIF\DT\zz=\varphi_\zz'(\Ee,\w,\zz)$
with $\Delta_{\bbM}^{-1}$ denoting the inverse linear operator to
$\Delta_{\bbM}={\rm div}(\bbM\nabla\cdot)$ is based on the total free energy
$\int_\Omega \psi(\Ee,\w,\zz,\nabla\w,\theta)\,\d x$ on the considered body
(denoted by $\Omega$) and the modified dissipation potential expressed in
terms of the rate $\DT\chi$ instead of $\mu$ as $\int_\Omega
\widetilde\zeta(\w,\theta;\EE,\ZJ\Ep,\DT\w,\DT\chi,\nabla\ZJ\Ep)\,\d x$
together with the total kinetic energy $\int_\Omega\frac\varrho2|\vv|^2\,\d x$.
The potential $\widetilde\zeta$ replaces the term
$\frac12\bbM|\nabla\mu|^2$ in \eqref{dissip-pot} by the
nonlocal term $\frac12\bbM|\nabla\Delta_{\bbM}^{-1}\DT\zz|^2$,
cf.\ \cite[Remark 7.6.4]{KruRou19MMCM}.
The system (\ref{eq}b--f) for a fixed temperature can then
be formally obtained by the Hamiltonian variational generalized
for nonconservative (dissipative) systems \cite{Bedf85HPCM}.
This is reflected by the correct energetics of the mechanical
part, as discussed below in Section~\ref{sec-termodynam}.
The quasistatic variant (i.e.\ with $\varrho=0$) has thus so-called
Biot's structure, cf.\ e.g.\ \cite[Sect.4]{Roub14NRSD}.

\section{Some specifications}\label{sec-specific}

An example of the stored energy occurring in 
\eqref{free} is
\begin{align}\nonumber
&\varphi(\Ee,\w,\zz)=\frac32K_\text{\sc e}^{}|{\rm sph}\Ee|^2\!
+G_\text{\sc e}^{}(\alpha)|{\rm dev}\Ee|^2\!+\frac1{2\kappa}G_\text{\sc h}\alpha^2
\\&\qquad\qquad\ \
+\frac 32K_\text{\sc b}^{}\Big|{\rm sph}\Ee -\frac{\zz{-}\phi}{3\beta}\bbI\Big|^2
\ \ \text{ with }\ \w=(\alpha,\phi)\,,
\label{stored}\end{align}
where $K_\text{\sc e}^{}$ is the bulk elastic modulus and
$G_\text{\sc e}^{}$ is the shear elastic modulus with the physical
dimension Pa\,=\,J/m$^3$, $\beta$ is the Biot coefficient,
and $K_\text{\sc b}^{}=\beta^2B$ with $B$ being the so-called Biot modulus.
With this $K_\text{\sc b}^{}$, the ``Biot'' term in \eqref{stored}
can equivalently be written as $\frac12B|\beta\,{\rm tr}\Ee{-}\zz{+}\phi|^2$.
The elastic stress ${\bm S}_\text{\sc e}^{}=\varphi_\Ee'(\Ee,\w,\zz)$ is now
\begin{align}
  {\bm S}_\text{\sc e}^{}=3(K_\text{\sc e}^{}{+}K_\text{\sc b}^{}){\rm sph}\Ee
  +2G_\text{\sc e}^{}(\alpha){\rm dev}\Ee-
  K_\text{\sc b}^{}\frac{\zz{-}\phi}{3\beta}\bbI
\label{Se}\end{align}
while the ``chemical'' potential $\mu=\varphi_\zz'(\Ee,\w,\zz)$ is now
a so-called {\it pore pressure}
\begin{align}
\mu=B(\zz-\phi-\beta\,{\rm tr}\Ee)\,.
\label{mu}\end{align}
The term $G_\text{\sc h}\alpha^2/(2\kappa)$ in \eqref{stored} reflects
the energy of damage -- microscopically, damage causes
some internal surfaces of voids in the material and
it just bears some additional energy. Values of the damage variable
$\alpha$ are standardly assumed to range over the interval $[0,1]$. 
Here we use the usual convention in geophysics that $\alpha=0$
means no damage while $\alpha=1$ means complete damage (in contrast
to engineering where the convention is opposite). The shear
modulus $G_\text{\sc e}^{}(\cdot)$ is naturally a decreasing function
of $\alpha$ and, assuming $G_\text{\sc e}'(1)=0$, the profile of $\alpha$
is kept within the interval $[0,1]$. The
$G_\text{\sc h}\alpha^2/(2\kappa)$ term is important because it yields
a driving thermodynamical force for {\it healing}, which means that $\alpha$ falls to 0
in unstressed (or only slightly stressed) rocks tempting
to minimize stored energy. This term also ``cooperates'' with the
gradient term in \eqref{free} in a way that, for $\kappa>0$ small,
the term $G_\text{\sc h}\alpha^2/(2\kappa)$ forces $\alpha$ to be small
in most spots of the domain $\Omega$ so that the damage can possibly
occur only on small regions (cracks).
Actually, this model is a popular approximation of sharp fracture, called a
{\it phase-field fracture}, advancing a static scalar study in
\cite{AmbTor90AFDP} towards vectorial dynamic cases, 
cf.\ \cite{KruRou19MMCM,Roub19MDDP} or,
for
application in modelling of tectonic fault rupture
or birth, \cite{RouVod19MMPF}.

As for the force (or, more precisely, acceleration) $\ff$ in \eqref{eq2},
it typically comes from the gravitational potential
as $\ff=-\nabla u$ with the gravitational potential
satisfying the equation 
\begin{align}\label{gravity}
  \frac1\Frakg\Delta u=\varrho
\end{align}
considered on the whole Universe $\R^3$ (with $\varrho=0$ outside the body
$\Omega$), where 
$\Frakg$ is the (normalized) gravitational constant, i.e.\
$\Frakg=4\pi\cdot6.673\times$Nm$^2$/kg$^2$.
In general, $\ff$ is rather non-potential if also
the centrifugal and Coriolis accelerations are included; then
$\ff=-\nabla u-2{\bm\omega}{\times}\vv
-{\bm\omega}{\times}({\bm\omega}{\times}\xx)$
with ${\bm\omega}$ a given angular velocity (as a vector)
and $\xx$ a position.

As for the dissipation potential $\zeta_1$ in \eqref{dissip-pot},
its natural form is 
\begin{align}
\zeta_1(\EE(\vv))=
\frac32K_{\rm v}|{\rm sph}\,\EE(\vv)|^2+G_{\rm v}|{\rm dev}\,\EE(\vv)|^2
 \label{dissip1-example}\end{align}
where $K_\text{\sc v}^{}$ is the viscous bulk modulus and
$G_\text{\sc v}^{}$ is the viscous shear modulus with the physical
dimension Pa\,s.

As for the dissipation potential $\zeta_2(\w,\theta;\cdot)$,
two basic options are positively homogeneous of degree one or two
corresponding to activated {\it plasticity} or Maxwellian {\it creep},
respectively. An interesting option is a combination of both these
phenomena. To this aim, one should consider two tensorial internal
variables, i.e.\ $\Ep=(\Ep_1,\Ep_2)$, and then
\begin{align}
\!\!\!\zeta_2(\w,\theta;\ZJ\Ep_1,\ZJ\Ep_2)
=G_{\text{\sc y}}^{}(\alpha,\theta)|\ZJ\Ep_1|
+\frac12G_\text{\sc m}^{}(\theta)|\ZJ\Ep_2|^2.\!\!
\label{dissip2-example}\end{align}
The additive decomposition of rates \eqref{eq3}
is then to be written as
\begin{align}
\ZJ\Ee+\ZJ\Ep_1+\ZJ\Ep_2=\EE(\vv)\,.
\end{align}
Notably, $\zeta_2(\w,\theta;\cdot,\cdot)$ is convex but
non-differentiable in terms of the component $\Ep_1$ at ${\bm 0}$.
Then, in fact, the equation \eqref{eq4} turns to be rather
an inclusion (or a variational inequality), which is nowadays
a standard concept from convex analysis; we omit such technicalities
here. This non-differentiability at $\Ep_1={\bm 0}$ allows
to model evolution of $\Ep_1$ as an activated process. More specifically, 
the yield-stress modulus $G_{\text{\sc y}}^{}=G_{\text{\sc y}}^{}(\alpha,\theta)$
in Pa\,=\,J/m$^3$ is an activation
stress threshold below which $\Ep_1$ does not evolve while, if
$|{\rm dev}{\bm S}_{\text{\sc e}}^{}|=G_{\text{\sc y}}^{}$,
this inelastic strain starts evolving --
the material (rock) {\it plastifies} and large
inelastic deformation may be accommodated 
fast along narrow shear bands - cataclasic zones
of tectonic faults. On the other hand, the
Maxwellian shear modulus $G_{\text{\sc m}}^{}$ is
typically quite large (about $10^{21\pm3}$Pa\,s
in the Earth crust or upper mantle) and this option facilitates
{\it creep} ({\it aseismic slip}) during long lasting
moderate shear stresses. The Maxwellian rheology is
also a basic model for seismic wave propagation
because of low attenuation possibility of propagation
of high-frequency waves. The combination of visco-elasto-plasticity
was used in geophysical modelling e.g.\ in
\cite{BabSob08HRNM,DiPrGe19SZUM,DDGP18SBMB,Gery19INGM,GerYue07RCMM,HeGeDi18IRSD,JacCac20MMBD,PopSob08TTDT,ReLYue03MSZG,ZWGM19MMES}. 

The flow-rule \eqref{eq5} for the scalar-valued internal variables
$\w=(\alpha,\chi)$ now takes a more specific form
\begin{align}\nonumber
  &\zeta_3\Big(\Big(\!\!\begin{array}{c}\alpha\\[-.2em]\phi\end{array}\!\!\Big),
  \theta;
  \Big(\!\!\begin{array}{c}\DT\alpha\\[-.2em]\DT\phi\end{array}\!\!\Big)\Big)
  \\[-.2em]&\qquad
  +\Big(\!\!\begin{array}{c}G_\text{\sc e}'(\alpha)|{\rm dev}\Ee|^2\!
  +G_\text{\sc h}\alpha/\kappa\\[-.1em] 
  B(\beta\,{\rm tr}\Ee-\zz+\phi)\end{array}\!\!\Big)
  =\Big(\!\!\begin{array}{c}\kappa\Delta\alpha
  \\[-.2em]\kappa\Delta\phi\end{array}\!\!\Big)\,.
  \label{dissip3-example}\end{align}
Typically, like $\zeta_2(\w,\theta;\cdot)$, the dissipation potential
$\zeta_3(\w,\theta;\cdot,\cdot)$ is convex
but non-differentiable at $(0,0)$. This non-differentiability serves for
modelling an {\it activation} character of evolution of these variables, meaning
that damage and porosity do not evolve unless there are enough driving
forces for it. In case of damage, it is reasonable to assume that 
the value $\lim_{\DT\alpha\to0+}[\zeta_3]_{\DT\alpha}'(\w,\theta;\DT\alpha,\DT\w)>0$
is positive, being called a
fracture toughness, while it is a reasonable modelling ansatz that
$\lim_{\DT\alpha\to0-}[\zeta_3]_{\DT\alpha}'(\w,\theta;\DT\alpha,\DT\w)=0$, which
allows for slow healing even when the driving force is small (or vanishes),
cf.\ \cite[Fig.\,2]{RoSoVo13MRLF}. Thanks to the mentioned non-smoothness, the
system of equations \eqref{dissip3-example} and thus also \eqref{eq5} are
actually variational inequalities. Let us still note that considering the
potential $\zeta_3$ acting simultaneously on the damage rate and porosity rate
allows for modelling a coupling between these two processes, as considered
(but only for a linear $[\zeta_3]_{\DT\w}'(\w,\theta,\cdot)$)
in \cite{HaLyAg04CEDP,LyaHam07DWFF} when writing \eqref{eq5} in an explicit
form as $\DT w=[[\zeta_3]_{\DT\w}'(\w,\theta;\cdot)]^{-1}
(\kappa\Delta\w-\varphi_\w'(\Ee,\w,\chi))$.
Notably, the porosity $\phi$ in this model \eqref{stored} evolves due to 
  the driving force is $-\mu$ and, in particular, is influenced
  by the spherical part of $\Ee$ while  
  the evolution of damage $\alpha$ is influenced by the deviatoric part of
  $\Ee$. Decreasing of porosity $\phi$ is called {\it compaction of rocks}.
  Keeping $\phi\ge0$ in this model can be ensured by increasing
  dissipation $\zeta_3(\alpha,\phi,\theta;\DT\alpha,\cdot)$ if $\phi\to0+$.

\begin{figure}
\psfrag{e}{\small $\EE(\uu)$}
\psfrag{a}{\small $\alpha$}
\psfrag{theta}{\small $\theta$}
\psfrag{w,theta}{\small $\alpha,\theta$}
\psfrag{devEe}{\small ${\rm dev}\Ee$}
\psfrag{sphEe}{\small ${\rm sph}\Ee$}
\psfrag{Ep1}{\small $\Ep_1$}
\psfrag{Ep2}{\small $\Ep_2$}
\psfrag{s}{\small $\sigma$}
\psfrag{e1}{\small ${\rm dev}\Ee+\Ep$}
\psfrag{ee2}{\small ${\rm sph}\,e_{\rm el}$}
\psfrag{G}{\small $G_\text{\sc e}^{}$}
\psfrag{Gkv}{\small $G_{_{\rm KV}}$}
\psfrag{K}{\small $K_\text{\sc e}^{}$}
\psfrag{Kkv}{\small $K_\text{\sc v}^{}$}
\psfrag{KB}{\small $K_\text{\sc b}^{}$}
\psfrag{Gkv}{\small $G_\text{\sc v}^{}$}
\psfrag{Gy}{\small $G_\text{\sc y}^{}$}
\psfrag{Gm}{\small $G_\text{\sc m}^{}$}
\psfrag{w}{\small $\mu=\mu(\Ee,\phi,\zz)$}
\psfrag{b}{\small $\beta$}
\psfrag{g}{\small $\ff$}
\psfrag{deviatoric}{\footnotesize\sf deviatoric part}
\psfrag{spherical}{\footnotesize\sf spherical part}
\begin{center}
\includegraphics[width=23em]{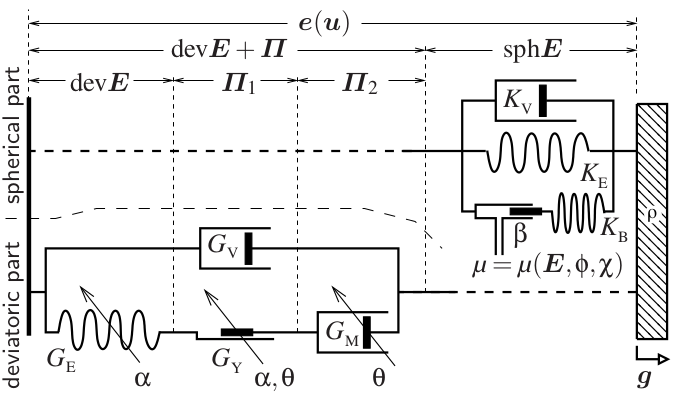}
\end{center}
\vspace*{-1.6em}
\caption{\sl Schematic diagram for the viscoelastic Jeffreys rheology
  which is subjected to damage $\alpha$ and depending on temperature
  in the deviatoric part 
  while the Kelvin-Voigt poro-visco-elastic rheology in 
  the spherical (volumetric) part with $\mu$ a pore pressure from \eqref{mu}.
}
\label{fig-mixed-response}
\end{figure}

  The parallel combination of Maxwellian viscoelastic rheology
and Stokes fluidic rheology is called {\it Jeffreys' rheology},
used in particular in \cite{LyHaBZ11NLVE}
and a combination with plasticity in \cite{RoSoVo13MRLF,RouTom??CMPE}.
The overall mechanical rheological model behind (\ref{eq}a--d) with 
\eqref{stored}, \eqref{dissip1-example}, and
\eqref{dissip2-example} is schematically depicted in
Figure~\ref{fig-mixed-response}. The part of the resting
model, i.e.\
the scalar internal-variables evolution and
fluid and heat transfer (\ref{eq}e--g), is not depicted in this
figure, however.

The philosophy of the anisothermal enhancement of Kelvin-Voigt/Jeffreys
viscoplastic model is to allow for, beside the friction model \eqref{RS-friction}
below, a mechanically enhanced
{\it Stefan-type phase transformation} a'la ice-water transformation,
as devised in \cite{Roub??SPTC}.
More specifically, when $G_\text{\sc m}^{}=0$, the elastic shear response
is completely suppressed
and the medium becomes {\it viscoelastic fluid}. This situation makes the
deviatoric elastic stress ${\rm dev}\bm{S}_\text{\sc e}^{}$ zero (so that
${\rm dev}\Ee=0$) and $\Ep_2$ freely evolving, which causes $\ZJ\Ep_2=\EE(\vv)$
so that $\Ep_2=\EE(\uu)-\Ep_1$. The Jeffreys solid rheology in the deviatoric
response then degenerates in the Stokes fluid, while the spherical response
remains as the Kelvin-Voigt solid, composing thus a viscoelastic fluid
(modelling magma, cf.\ Section~\ref{sec-PT} below). This allows for propagation
of longitudinal waves in fluidic regions while shear waves are suppressed (or
transformed and reflected on the boundaries). Importantly, ${\rm dev}\Ee=0$
means that $\Ee=p\bbI/K_\text{\sc e}^{}$ for some hydrostatic pressure 
and \eqref{eq3} reduces to $\DT p=K_\text{\sc e}^{}{\rm div}\,\vv$;
here one exploits \eqref{ZJ-property}.
Also, controlling the modulus $G_\text{\sc m}^{}$ by temperature
then allows to model melting of rocks to magma and, conversely, solidification
of magma to solid rocks, cf.\ Section~\ref{sec-PT} below for a more detailed
discussion. Notably, although the inelastic strain $\Ep_2$ does not
influence the responses in the fluidic state, it nevertheless continues
evolving with $\EE(\vv)$ so that the possible later solidification performs in
the actual configuration reflected by evolved $\Ep=\Ep_1{+}\Ep_2$ while the
former configuration at the melting time is naturally forgotten. This allows
for repeating arbitrarily many times solid-to-fluid or fluid-to-solid
transformations.

Let us still note that the vector of scalar internal variables
$\w$ opens possibilities for expansions or modifications of the
presented model. An example is a combination of damage with so-called
breakage \cite{LyaBeZ14CDBF,LyaBeZ14DBRM,LBIM16DRDB} for solid-granular
materials, where again the coupling with damage as in \eqref{dissip3-example}
is considered. Moreover, some coefficients in the dissipative part of the
model can be easily made dependent more generally on other variables without
changing the structure of the resulting equations and the calculations
in Section~\ref{sec-termodynam} below, i.e.\
$K_\text{\sc v}^{}=K_\text{\sc v}^{}({\rm tr}\Ee,\w,\zz,\theta)$,
$G_\text{\sc v}^{}=G_\text{\sc v}^{}({\rm tr}\Ee,\w,\zz,\theta)$,
$G_\text{\sc m}^{}=G_\text{\sc m}^{}({\rm tr}\Ee,\w,\zz,\theta)$,
$\bbK=\bbK({\rm tr}\Ee,\w,\zz,\theta)$, or
$\bbM=\bbM({\rm tr}\Ee,\w,\zz,\theta)$.
In particular, the dependence of $G_\text{\sc v}^{}$ on ${\rm tr}\Ee$ (i.e.\ on
the pressure $K_\text{\sc e}^{}{\rm tr}\Ee$) and on $\zz$ allows for
making the temperature of the solid-fluid transformation pressure and
water-content dependent, which is indeed an important phenomenon in Earth
mantle.
\nopagebreak
\section{Tectonic earthquake modelling}\label{sec-EQ}
In contrast to engineering fracture models, geophysical
fast fracture is to be followed by slow healing (also called
{\it aging}). This prepares a condition for another fracture and thus,
considering application to tectonic faults, for re-occurring earthquakes.
It seems to bring special demands on successful computation
simulations to avoid tendency for aseismic slip instead of
a desired stick and healing after each rupture. There
exist two possibilities in literature, either to impose some instability
by making the stored energy 
nonconvex in terms of $\Ee$ when the rock is enough damaged or to impose
some slip velocity weakening (modelled here as some yield stress
$G_{\text{\sc y}}^{}$ weakening).

As to the first option, an ansatz which is 2-homogeneous in terms
of $\Ee$, suggested in \cite{LyaMya84BECS}, modifies \eqref{stored} as
\begin{align}\nonumber
&\varphi(\Ee,\w,\zz)=\frac32K_\text{\sc e}^{}(\alpha)|{\rm sph}\Ee|^2\!
+G_\text{\sc e}^{}(\alpha)|{\rm dev}\Ee|^2\!
\\[-.2em]&\nonumber\qquad\qquad\quad
-N_\text{\sc e}^{}(\alpha){\rm tr}\Ee\,|\Ee|+\frac1{2\kappa} G_\text{\sc h}\alpha^2
\\[-.2em]&\qquad\qquad\qquad\qquad
+\frac 32K_\text{\sc b}^{}\Big|{\rm sph}\Ee -\frac{\zz{-}\phi}{3\beta}\bbI\Big|^2
\label{Vladimir}\end{align}
with some non-Hookean modulus $N_\text{\sc e}^{}$ as an increasing
function of damage $\alpha$ which, for $\alpha\to0$, becomes
sufficiently large to make $\varphi(\cdot,\w,\zz)$ nonconvex.
This model was used in a series of articles,  cf.\ e.g.\
\cite{BeZi08CBEF,FMGA13SBFN,HaLyAg04CEDP,HLSD09BDDI,LyBZAg05VDRR,LBIM16DRDB,LHAB09NDRW,LyHaBZ11NLVE,LyaBeZ14DBRM}
and references therein.

The second option uses (as some departure point) the
{\it rate-and-state}-dependent Dieterich-Ruina {\it friction}
model, devised originally in an isothermal variant
as a frictional contact problem with some slip velocity
weakening of the friction coefficient
(called also a sliding resistance) in
\cite{Diet79MRF,Ruin83SISVFL}. This original model seems however 
incompatible with the 2nd law of thermodynamics (i.e.\ the entropy
Clausius-Duhem inequality), as articulated in \cite[Sect.\,2]{Roub14NRSD}. 
Yet, a similar instability effect can be obtain by replacing
velocity by temperature while relying on that local temperature
variation is well related with slip rate during fast
earthquakes. This reflects the phenomenon that temperature
can locally rise by many hundreds degrees. Such intense heat production during
strong earthquakes is referred to as a {\it flash heating},
cf.\ \cite{BeTuGo08CRPB,Bizz12MLFI,BizCoc03SWBD,Rice99FHAC,RicCoc06SFRE}.
``Translating'' the contact model to the bulk model as suggested in
\cite[Sect.\,6.3]{Roub14NRSD} can be performed by making the yield stress
$G_\text{\sc y}^{}$ dependent on damage (aging) and temperature as
\begin{align}
  G_\text{\sc y}^{}(\alpha,\theta)=G_\text{\sc y0}^{}
  +a\Big({\rm ln}\frac{\theta}{v_{\rm ref}\!}+1\Big)
  +b\Big({\rm ln}\frac{\!v_{\rm ref}^{}\alpha\!}{d_{\rm c}}+1\Big)\
\label{RS-friction}\end{align}
with (given) parameters $a$ and $b$ (called respectively the direct-effect
and evolution friction parameters), $d_{\rm c}$ a characteristic
slip memory length, and $v_{\rm ref}$ a reference velocity.
The desired velocity weakening occurs if $a<b$.

A simplified friction model with only one parameter 
is sometimes also considered under the name rate-dependent friction
\cite{Diet79MRF,LyBZAg05VDRR,Ruin83SISVFL,TonLav18SSTE}
and was analyzed in \cite{Miel18TECI} as far as its stability.

Actually, earthquakes on faults in hydrated rocks can be
influenced by the water content by causing additional weakening
especially under remarkable flash heating, cf.\ e.g.\
\cite{Ches95RMWC,Lauc80FHFP,Rice06HWFE}
or Sect.\,3.1 in \cite{PGYD20SHMM}. This phenomenon can be reflected in our
model directly by making $G_\text{\sc y}^{}$ dependent also
on the water content $\chi$.

\section{Rock--magma transition modelling}\label{sec-PT}
The ability of the model to cover solid-liquid phase transformation
mentioned already in Section~\,\ref{sec-specific} can be exploited 
to model {\it melting\,/\,solidification} within
transformation between solid rocks and fluidic magma.
The specific feature is that magma is not exactly a liquid
(in particular not an ideal Stokes fluid), being reported as a
partially melted rock where shear waves can still propagate although
with a reduced velocity \cite{Schm04V}.

Another feature is that the melting\,/\,solidification
does not occur at a single melting temperature (in contrast e.g.\ to a
single-component ice-water phase transition) because rocks contain many
constituents with different melting temperatures \cite{Schm04V}.
In particular, the multi-component rock-magma
transition is rather ``fuzzy'' in some temperature
interval and, instead of a sharp interface, a {\it mush zone} is expected.
This can easily be reflected by considering $\gamma$ in \eqref{eq7}
continuous (only with a bigger slope in melting temperature interval).

When melting is completed, $G_\text{\sc m}^{}$ in \eqref{dissip2-example}
``nearly'' (but not completely) vanishes.
Although $G_\text{\sc m}^{}$ remains non-zero in magma, typical values
$10^{4\pm3}$ Pa\,s are by many orders
lower that in rocks where $G_\text{\sc m}^{}$ is about $10^{21\pm3}$Pa\,s,
as already mentioned. Thus, from longer time scales, magma is not
far from being quite ideal fluid.

Simultaneously, when making $[\zeta_3]_{\w}'(\w,\theta;\cdot)$ vanishing
for $\theta$ above melting temperature, both the damage $\alpha$
and porosity $\phi$ fall fast to zero after the
compressed rock is melted to compressed liquid magma.

There are other phenomena \cite{Schm04V} which can be incorporated
in the model. Specifically, viscosity of magmas slightly decreases
with increasing pressure, which suggests to make $G_\text{\sc m}^{}$
dependent also on ${\rm tr}\Ee$, and decreases with the shear
rate, i.e.\ magma is rather a non-Newtonian viscoelastic fluid.  
This suggests to use \eqref{dissip1-example} 
with the exponent $p<2$ instead of 2. The latter variant is called a
shear thinning. Other phenomenon in the rock-magma system which can be
potentially covered by this model is 
{\it volcanic earthquakes} arising from shear fracture in rocks adjacent to
magma chambers.

A general phenomenon of {\it buoyancy} in the whole mantle \cite{ScTuOl01MCEP}
is particularly articulated in volcanism because magma is lighter than rocks
especially below Mohorovi\v ci\v c discontinuity \cite{Schm04V}. Buoyancy
is then an important phenomenon which drives movement of magma in mantle and
eventually may lead to volcanic activity on the Earth surface.
The buoyancy effects can be naturally
modelled by {\it thermal expansion} of the material. In particular, above
melting temperature the material substantially expands and decreases the density
$\varrho$, which then decreases the force $\varrho\ff$ on the right-hand side
of \eqref{eq2} and thus lead to the buoyancy effect.
The term $\frac32K_\text{\sc e}^{}|{\rm sph}\Ee|^2$ in
\eqref{stored} is to be replaced by
$\frac32K_\text{\sc e}^{}|{\rm sph}\Ee-\varepsilon(\theta)\mathbb I|^2$
with some function $\varepsilon(\cdot)$ which increases in particular when
temperature $\theta$ passes through the interval of melting temperatures. In
other words, the free energy \eqref{free} augments by a mixed term
$-3K_\text{\sc e}^{}\varepsilon(\theta){\rm tr}\Ee$ while the function
$\HEAT(\cdot)$ in \eqref{free} contains now also
$\frac12K_\text{\sc e}^{}\varepsilon^2(\cdot)$, i.e. now
\begin{align}\nonumber
\psi(\Ee,\w,\zz,\nabla\w,\theta)&=\varphi(\Ee,\w,\zz)+\frac\kappa2|\nabla\w|^2
\\&\ \ \ \
+\HEAT(\theta)-3K_\text{\sc e}^{}\varepsilon(\theta){\rm tr}\Ee\,.
\label{free+}\end{align}
The elastic stress ${\bm S}_\text{\sc e}^{}$ in \eqref{Se} 
then augments by a pressure stress
$-3K_\text{\sc e}^{}\varepsilon(\theta)\mathbb I$ while the heat-transfer
equation \eqref{eq7} augments by an adiabatic heat source/sink
$3\theta K_\text{\sc e}^{}\varepsilon'(\theta){\rm tr}\DT\Ee$ and
uses $\gamma=\gamma(\Ee,\theta)$ depending now also on ${\rm sph}\Ee$ as
\begin{align}
\gamma(\Ee,\theta)=h(\theta)-\theta h'(\theta)-3K_\text{\sc e}^{}
\big(\varepsilon(\theta){-}\theta\varepsilon'(\theta)\big){\rm tr}\Ee\,;
\label{gamma+}\end{align}
here we assume, without loss of generality, that
$\varepsilon(0)=0$, cf.\ \cite[Remark 8.1.4]{KruRou19MMCM}.
Thus the heat capacity is now $c({\rm tr}\Ee,\theta)=-\theta(\HEAT''(\theta)
{-}3K_\text{\sc e}^{}\varepsilon''(\theta){\rm tr}\Ee)$
and, in general, depends also on the pressure through ${\rm tr}\Ee$
unless $\varepsilon(\cdot)$ is affine.
 
Yet, it should be said that there are some other mechanisms which may contribute
to magma generation than mere increase of temperature.

\section{Thermodynamics of the model (7) 
}\label{sec-termodynam}

Let us derive the energetics behind the system \eqref{eq}. To this goal, we
must consider some domain, let us denote it by $\Omega\subset\R^3$, and its
boundary $\Gamma$, and to consider some boundary conditions. For simplicity,
let us consider
\begin{subequations}\label{BC}\begin{align}\label{BC1}
    &\!\!\!\!\vv{\cdot}\nn=0,\ \,\big[\big(\bm{S}_\text{\sc e}^{}{+}{\bm S}_\text{\sc v}^{}
         {+}{\bm S}_{\rm str}\big)\nn\big]_\text{\sc t}^{}={\bm t},\ \,(\nn{\cdot}\nabla)\Ep=0,
\\&\!\!\!\!(\nn{\cdot}\nabla)\w=0,\ \ \ \ 
         \nn{\cdot}\nabla\mu=0,\ \ \ \ \nn{\cdot}\bbK\nabla\theta=j_{\rm ext}
\label{BC2}\end{align}\end{subequations}
with $\nn$ denoting the outward unit normal
to $\Gamma$, $[\cdot]_\text{\sc t}^{}$ the tangential component of a vector on $\Gamma$,
and ${\bm t}$ a traction force
and $j_{\rm ext}$ an external heat flux through $\Gamma$.

Multiplying the continuity equation \eqref{eq1} by $|\vv|^2/2$, we obtain
\begin{align}\nonumber
 \pdt{}\Big(\frac\rho2|\vv|^2\Big)&=\rho\vv{\cdot}\pdt\vv
  +\pdt\rho\frac{|\vv|^2}2
\\&
  =\rho\vv{\cdot}\pdt\vv
-(\vv{\cdot}\nabla\rho)\frac{|\vv|^2}2-\rho({\rm div}\,\vv)\frac{|\vv|^2}2\,.
\label{formula0}\end{align}

The momentum equation \eqref{eq2} should be multiplied by the velocity $\vv$.
The convective term $\varrho(\vv{\cdot}\nabla)\vv{\cdot}\vv$ contained in
$\varrho\DT\vv{\cdot}\vv$ gives, when integrated over $\Omega$ and processed by Green's
formula, that
\begin{align}\nonumber
&\int_\Omega\rho(\vv{\cdot}\nabla)\vv{\cdot}\vv\,\d x
=-\int_\Omega\vv{\cdot}{\rm div}(\rho\vv{\otimes}\vv)\,\d x
\\[-.2em]&\ \nonumber
=-\int_\Omega(\vv{\cdot}\nabla\rho)|\vv|^2
+\rho(\vv{\cdot}\nabla)\vv{\cdot}\vv+\rho({\rm div}\,\vv)|\vv|^2\,\d x
\\[-.5em]&\nonumber
\qquad\qquad
=-\int_\Omega(\vv{\cdot}\nabla\rho)\frac{|\vv|^2}2+\rho({\rm div}\,\vv)\frac{|\vv|^2}2\,\d x
\\[-.2em]&\qquad\qquad\qquad\quad
=\frac{\d}{\d t}
\int_\Omega\frac\rho2|\vv|^2\,\d x-\int_\Omega\rho\pdt\vv{\cdot}\vv\,\d x\,.
\label{formula1}\end{align}

The flow rule \eqref{eq4} for $\Ep$ is to be multiplied by $\ZJ\Ep$. Using
\eqref{eq3} and the matrix algebra $A{:}(BC)=(B^\top\!A){:}C=(AC^\top){:}B$,
from the term ${\rm dev}\bm{S}_\text{\sc e}^{}$ we obtain
\begin{align}\nonumber
  &{\rm dev}\bm{S}_\text{\sc e}^{}{:}\ZJ\Ep=
\bm{S}_\text{\sc e}^{}{:}\ZJ\Ep=\bm{S}_\text{\sc e}^{}{:}\EE(\vv)-\bm{S}_\text{\sc e}^{}{:}\ZJ\Ee
\\&\qquad\nonumber
=\bm{S}_\text{\sc e}^{}{:}\EE(\vv)-
\varphi_\Ee'(\Ee,\w,\zz){:}\Big(\pdt\Ee+(\vv{\cdot}\nabla)\Ee\Big)
\\&\nonumber\qquad\qquad
-\bm{S}_\text{\sc e}^{}{:}\big(\Ee\,{\rm skew}(\nabla\vv)
-{\rm skew}(\nabla\vv)\Ee\big)
\\&\qquad\nonumber
=\Big(\bm{S}_\text{\sc e}^{}+{\rm skew}\big(\bm{S}_\text{\sc e}^{}\Ee^\top\!\!
-\bm{S}_\text{\sc e}^\top\Ee\big)\Big){:}\EE(\vv)
\\[-.3em]&\qquad\qquad\nonumber
-\varphi_\Ee'(\Ee,\w,\zz){:}\Big(\pdt\Ee+(\vv{\cdot}\nabla)\Ee\Big)
\\&\qquad
=\bm{S}_\text{\sc e}^{}{:}\EE(\vv)
-\varphi_\Ee'(\Ee,\w,\zz){:}\Big(\pdt\Ee+(\vv{\cdot}\nabla)\Ee\Big)\,.
\label{formula2}\end{align}
Let us note that  actually
$\bm{S}_\text{\sc e}^{}\Ee^\top\!\!-\bm{S}_\text{\sc e}^\top\Ee=0$ because
the transport by corotational derivatives keeps elastic strain and stress tensors
symmetric, and 
thus the stress in \eqref{eq2} is symmetric. This is consistent with the
expected symmetry of the Cauchy stress except when also some dipoles
(as magnetization or polarization) were transported, cf.\ \cite{Roub??TMPE}.

The flow rule \eqref{eq5} for $\w$ is to be multiplied by $\DT\w$, which
gives rise in particular to the term
\begin{align}
  \varphi_\w'(\Ee,\w,\zz)\DT\w&=\varphi_\w'(\Ee,\w,\zz)\pdt\w
  +\varphi_\w'(\Ee,\w,\zz)(\vv{\cdot}\nabla)\w\,.
\label{formula3}\end{align}
Moreover, the gradient term in \eqref{eq5} can be handled by
the calculus together with the Green formula over $\Omega$ as
\begin{align}\nonumber
  &\!\!\!\!\!\!\int_\Omega\kappa\Delta\w{\cdot}\DT\w\,\d x=
\int_\Omega\kappa\Delta\w{\cdot}(\vv{\cdot}\nabla)\w\,\d x
-\frac{\d}{\d t}\int_\Omega\frac\kappa2|\nabla\w|^2\,\d x
\\[-.2em]&\nonumber
=\int_\Gamma\kappa(\nn{\cdot}\nabla)\w{\cdot}(\vv{\cdot}\nabla)\w\,\d S
-\frac{\d}{\d t}\int_\Omega\frac\kappa2|\nabla\w|^2\,\d x
\\[-.2em]&\nonumber\hspace{3em}
-\int_\Omega\Big(\kappa(\nabla\w{\otimes}\nabla\w){:}\EE(\vv)
-\frac\kappa2|\nabla\w|^2{\rm div}\,\vv\Big)\,\d x
\\&\nonumber
=\int_\Omega\Big(\frac\kappa2|\nabla\w|^2\bbI-
\kappa(\nabla\w{\otimes}\nabla\w)\Big){:}\EE(\vv)\,\d x
\\[-.5em]&\hspace{12em}
-\frac{\d}{\d t}\int_\Omega\frac\kappa2|\nabla\w|^2\,\d x\,.
\end{align}
We used the boundary condition $(\nn{\cdot}\nabla)\w=0$.
Here we can identify the Korteweg-like contribution
$\frac\kappa2|\nabla\w|^2\bbI-\kappa(\nabla\w{\otimes}\nabla\w)$
in the structural stress $\bm{S}_{\rm str}$ in \eqref{eq2}.

The product of the two equations in \eqref{eq6} gives
\begin{align}
  \varphi_\zz'(\Ee,\w,\zz)\pdt\zz+\DIF\DT\zz^2
&=\mu\:{\rm div}(\bbM\nabla\mu)
  -\varphi_\zz'(\Ee,\w,\zz)\vv{\cdot}\nabla\zz.
\label{formula4}\end{align}
Integrating $\mu\:{\rm div}(\bbM\nabla\mu)$ over $\Omega$ and
using Green's formula gives
the dissipation rate due to diffusion $\int_\Omega\bbM|\nabla\mu|^2\,\d x$.

Summing the partial time-derivative terms in  \eqref{formula2},
\eqref{formula3}, and \eqref{formula4}, we can use the calculus
\begin{align}\nonumber
&\varphi_\Ee'(\Ee,\w,\zz){:}\pdt\Ee
  +\varphi_\w'(\Ee,\w,\zz){\cdot}\pdt\w
  \\&\qquad\qquad\qquad\quad
  +\varphi_\zz'(\Ee,\w,\zz)\pdt\zz
  =\pdt{}\varphi(\Ee,\w,\zz)
\end{align}
to see the main part of the stored energy rate,
cf.\ \eqref{energy+} and  \eqref{energy+++} below.
Furthermore, 
summing the convective terms in \eqref{formula2},
\eqref{formula3}, and \eqref{formula4} and using the
Green formula when integrating them over $\Omega$,
we obtain
\begin{align}\nonumber
  &\int_\Omega\Big(
  \varphi_\Ee'(\Ee,\w,\zz){:}(\vv{\cdot}\nabla)\Ee
\\[-.3em]&\nonumber\qquad
  +\varphi_\w'(\Ee,\w,\zz)(\vv{\cdot}\nabla)\w
+\varphi_\zz'(\Ee,\w,\zz)\vv{\cdot}\nabla\zz\Big)\,\d x
\\&\nonumber
=\int_\Omega\nabla\varphi(\Ee,\w,\zz){\cdot}\vv\,\d x
\\&\nonumber
=\int_\Gamma\varphi(\Ee,\w,\zz)\vv{\cdot}\nn\,\d S
-\int_\Omega\varphi(\Ee,\w,\zz){\rm div}\,\vv\,\d x
\\&
=-\int_\Omega\varphi(\Ee,\w,\zz)\bbI{:}\EE(\vv)\,\d x\,;
\end{align}
here we used the boundary condition $\vv{\cdot}\nn=0$
from \eqref{BC1} and obtained one part of the pressure
contribution $\varphi(\Ee,\w,\zz)\bbI$ in the structural
stress $\bm{S}_{\rm str}$ in \eqref{eq2}.

Altogether, summing the above mentioned tests of (\ref{eq}a,c,d,e)
and using the boundary conditions in \eqref{BC}, we can exploit the
suitably chosen structural stress $\bm{S}_{\rm str}$ in \eqref{eq2}. We can
also see cancellation of the right-hand side in
\eqref{formula1} with two last terms in 
\eqref{formula0} integrated over $\Omega$.
Thus we eventually obtain the desired energy balance
\begin{align}\nonumber
  &\frac{\d}{\d t}
  \int_\varOmega\!\!\!\!\!\linesunder{\frac\varrho2|\vv|^2\!\!}{kinetic}{$^{^{}}$energy}
  \!\!\!\!\!\!\!+\!\!\!\!\!\!\!\lineunder{\varphi(\Ee,\w,\zz)
    +\frac{\kappa}2|\nabla\w|^2\!\!}{stored
    energy}\!\!\!\!\d x
  \\[-.2em]&\nonumber\ \qquad\qquad
  +
  \int_\varOmega\!\!\!\!\!\!\lineunder{\!\!
    \xi(\w,\theta;\EE(\vv),\ZJ\Ep,\DT\w,\nabla\mu,\DT\zz,\nabla\ZJ\Ep)
    _{_{_{_{_{_{}}}}}}\!\!
  _{_{_{}}}\!}{dissipation rate}
  \!\!\!\!\!\!\d x
\\[-.2em]&\hspace*{1em}
=\int_\varOmega\!\!\!\!\!\!\!\linesunder{\varrho
  \ff{\cdot}\vv-\HEAT(\theta){\rm div}\,\vv_{_{_{_{}}}}\!\!}{power of external load}{and of adiabatic effects}\!\!\!\!\!\d x
  +\int_\varGamma\!\!\!\!\!\!\!\!\!\!\!\!\!\!\linesunder{\,{\bm t}{\cdot}\vv_{_{_{}}}\!}{power of}{traction load}\!\!\!\!\!\!\!\!\!\!\!\!\!\!\d S
  \,.
  \label{energy+}\end{align}

Adding \eqref{eq7} integrated over $\Omega$, the dissipation term $\xi$ and the
adiabatic terms on the right-hand side of \eqref{eq7} cancel. Using also the
last boundary condition in \eqref{BC2}, we thus obtain the
{\it total energy balance} 
\begin{align}\nonumber
  &\frac{\d}{\d t}
  \int_\varOmega
\!\!\!\!\!\linesunder{\frac\varrho2|\vv|^2\!\!}{kinetic}{$^{^{}}$energy}\!\!\!\!\!\!
  +\!\!\!\!\!\!\linesunder{\varphi(\Ee,
    \w,\zz)
    +\frac{\kappa}2|\nabla\w|^2\!\!}{stored}{$^{^{}}$energy}\!\!\!\!
+\!\!\!\!\!\!\!\!\!\!\!\!\!\!\!\!\!\!\!\!\!\!\!\linesunder{
   \W_{_{_{_{_{}}}}}\!\!}{internal heat}{$^{^{}}$energy (enthalpy)}
\!\!\!\!\!\!\!\!\!\!\!\!\!\!\!\!\!\!\!\!\!
\,\d x
\\&
\ \ =\ \ \int_\varOmega\!\!\!\!\!\!\!\!\!\!\!\!\!\!\!\!\!\!\!\linesunder{\varrho\ff{\cdot}\vv_{_{_{_{}}}}\!\!\!}{
    external mechanical}{$^{^{^{}}}$bulk power}\!\!\!\!\!\!\!\!\!+
  \!\!\!\!\!\linesunder{\!\!r\!\!_{_{_{_{_{}}}}}}{bulk}{heat power}\!\!\!\!\!\!\!\!\!\d x
  +\int_\varGamma\!\!\!\!\!\!\!\!\!\!\!\!\!\!\!\!\!\!\linesunder{
    {{\bm t}{\cdot}\vv_{_{_{}}}+j_{\rm ext}}_{_{}}\!\!\!\!\!\!}{power of traction and
  }{of
    external heating}\!\!\!\!\!\!\!\!\!\!\!\!\!\d S\,.
  \label{energy+++}\end{align}

For the case of a gravitation acceleration $\ff=-\nabla u$ with the potential
$u$ from \eqref{gravity}, we can be more specific as far as the external load
in \eqref{energy+} and \eqref{energy+++} concerns. Then the bulk force
$\varrho\ff$ in the momentum equation tested by $\vv$ together with
\eqref{gravity} tested by $\pdt{}u$ results to
\begin{align}\nonumber
  &\int_\Omega\!\!-\varrho\nabla u{\cdot}\vv\,\d x
  =\int_\Omega{\rm div}(\varrho\vv)u\,\d x
  \\\nonumber&\quad
  =-\int_\Omega\pdt\varrho u\,\d x
 =\int_\Omega\varrho\pdt u\,\d x-\frac{\d}{\d t}\int_\Omega\varrho u\,\d x
 \\&\qquad\
 =-\frac{\d}{\d t}\Big(\int_{\R^3}\frac1{2\Frakg}|\nabla u|^2\,\d x
  +\int_\Omega\varrho u\,\d x\Big)\,,
\end{align}
where we used also the continuity equation \eqref{eq1} and the boundary
condition $\vv{\cdot}\nn=0$. It reveals the energy of the gravitation
field and the energy of the mass $\varrho$ in this (self-induced) gravitation field.
The energy identities \eqref{energy+} and \eqref{energy+++} can then
be specified accordingly. In particular, if the system is isolated
in the sense ${\bm t}={\bm 0}$ and $j_{\rm ext}=0$, then \eqref{energy+++}
gives the {\it total energy conservation}
\begin{align}\nonumber
\int_\varOmega
\frac\varrho2|\vv|^2+\varphi(\Ee,\w,\zz)+\frac{\kappa}2|\nabla\w|^2
+\W+\varrho u\,\d x\ \ 
\\
+\int_{\R^3}\frac1{2\Frakg}|\nabla u|^2\,\d x={\rm const.}
\end{align}

An important attribute of the model, beside keeping the energetics
\eqref{energy+}--\eqref{energy+++}, is the entropy imbalance, i.e.\ 
the {\it Clausius-Duhem inequality}. The specific 
entropy $\eta=-\psi_\theta'=-\HEAT'(\theta)$ as an extensive variable
(in JK$^{-1}$m$^{-3}$) and its transport and production is governed by the
{\it entropy equation}: 
\begin{align}
  \pdt\eta+{\rm div}(\vv\,\eta)
=\frac{\xi-{\rm div}\,{\bm j}+r}\theta
\label{entropy-eq}\end{align}
with $\xi$ denoting the heat production rate by mechanical/diffusion
dissipative processes, cf.\ \eqref{dissip-rate},
and ${\bm j}$ the heat flux which is here governed by the Fourier law ${\bm j}=-
\bbK\nabla\theta$, cf.\ \eqref{eq7}. The ultimate assumptions $\xi+r\ge0$
and $\bbK$ positive then ensure the {\it entropy imbalance}
\begin{align}
\frac{\d}{\d t}\int_\Omega\eta\,\d x\ge
\int_\Gamma\!\!\!\!\!\lineunder{(\vv\eta-{\bm j}/\theta)}{entropy flux}\!\!\!\!\!\!\!\cdot\nn\,\d S
=\int_\Gamma\frac{j_{\rm ext}}\theta\,\d S
\label{entropy>}\end{align}
by the usual calculus using also the boundary conditions \eqref{BC},
relying on positivity of temperature.
Substituting $\eta=-\HEAT'(\theta)$ into \eqref{entropy-eq}, we obtain
the heat-transfer equation 
\begin{align}\nonumber
&c(\theta)\DT\theta
=\xi-{\rm div}\,{\bm j}-\W\eta\,{\rm div}\,\vv+r
\\&
\qquad\text{ with the heat capacity }\
c(\theta)=-\theta\HEAT''(\theta)\,;
\label{heat-eq-}\end{align}
note that temperature (in K) is an intensive variable and is thus transported
by the material derivative while the adiabatic heat 
term $\theta\eta\,{\rm div}\,\vv$ occurs on the right-hand side due to
the compressibility of the material.
Furthermore, the {\it internal energy} is given by the Gibbs relation
$\psi+\theta\eta$. In view of \eqref{free}, it splits here into the purely
mechanical part
$\varphi(\Ee,\w,\zz)+\frac\kappa2|\nabla\w|^2$  
and the
thermal part $\W=\HEAT(\theta)-\theta\HEAT'(\theta)=:\gamma(\theta)$.
The thermal internal energy $\W$ in Jm$^{-3}$ is again an extensive
variable and is transported like \eqref{entropy-eq}, resulting here to the
equation
\begin{align}
  \pdt\W+{\rm div}\big(\vv\,\W\big)=\xi-{\rm div}\,{\bm j}
  +\HEAT(\theta){\rm div}\,\vv+r\,,
\label{heat-eq}\end{align}
which reveals the structure of \eqref{eq7}.

For \eqref{entropy>}, we actually needed that $\theta$ is
positive, i.e.\ the model complies with the Nernst 3rd law of
thermodynamics. This is really guaranteed in the model
\eqref{eq} under some natural assumptions, namely
that $\gamma$ is increasing function of temperature with $\gamma(0)=0$,
and $h$ in \eqref{free} is concave with $h(0)=0$.
Then $\theta\ge0$ provided the initial temperature is non-negative and
provided the boundary heat flux $j_{\rm ext}$ in \eqref{BC2} is non-negative
or, more generally, $j_{\rm ext}$ is a continuous function of temperature
and is non-negative for $\theta=0$.

For the generalization \eqref{free+}, the entropy depends on ${\rm tr}\Ee$ as
$\eta=\eta({\rm tr}\Ee,\theta)=3K_\text{\sc e}^{}
\varepsilon'(\theta){\rm tr}\Ee-h'(\theta)$.
The right-hand side
of the mechanical energy balance \eqref{energy+} then expands by the
adiabatic term $-3K_\text{\sc e}^{}\varepsilon(\theta){\rm div}\,\vv$
while \eqref{energy+++} uses $\W=\gamma(\Ee,\theta)$ from \eqref{gamma+}.

\section{A semi-compressible variant}\label{sec-semi}

The core of the model, i.e.\ the mass and momentum
conservation (\ref{eq}a,b), is ``borrowed'' from fully
compressible fluids covering also gases. Most solid and liquid
materials (and in particular rocks and magmas) are not much compressible,
however. In particular, density variations are not much pronounced.
In a lot of studies, e.g.\
\cite{HaLyAg04CEDP,LBIM16DRDB,LHAB09NDRW,LyHaBZ11NLVE,Dint13SCST},
the mass density $\varrho$ is simply considered constant.
This would be well eligible in incompressible models in media
which are spatially homogeneous as far as mass density $\varrho$
concerns. Yet, incompressible models do not facilitate propagation
of longitudinal seismic waves, which would be an essential drawback
of such models in geophysical context. For this reasons, compressible
models can be considered in a simplified, compromising, so-called
{\it semi-compressible} \cite{Roub20QSF} variant relying on that
rocks and magmas are only
slightly compressible so that, assuming also that the particular
space-time region of interest in the mantle is relatively small and thus the
mass density $\varrho$ does not vary substantially.
 This is in particular a relevant simplification if pressure variations
are negligible comparing to the bulk elastic modulus.  In some
approximation, $\varrho$ can then be considered constant and the continuity
equation \eqref{eq1} is avoided. Yet, to keep energy conservation,
one has to involve a certain {\it structural acceleration} $\ff_{\rm str}$ into
the momentum equation \eqref{eq2}, which then looks as
\begin{align}\nonumber
  \!\!\!\varrho\DT\vv={\rm div}\big(\bm{S}_\text{\sc e}^{}
+{\bm S}_\text{\sc v}^{}+{\bm S}_{\rm str}\big)
+\varrho(\ff+\ff_{\rm str})
\quad\ \ \
\\[-.2em]
\text{ with }\ \ \ff_{\rm str}=
-\,\frac12({\rm div}\,\vv)\,\vv
\,.
\label{eq2+}\end{align}
The force $\varrho\ff_{\rm str}$ was invented in
\cite{Tema69ASEN}
rather for numerical purposes to approximate an incompressible model, and
recently physically analyzed in
\cite{Toma21ITST} where it was pointed out that this force does not comply
with Galilean invariancy principle. Of course, in only slightly compressible
materials, ${\rm div}\,\vv$ and also this extra force is only small
and the Galilean invariancy is violated only a little,
which is an acceptable price for keeping the energetics holding in this
simplified model.

The calculus \eqref{formula1} is to be then replaced by
\begin{align}\nonumber
&\int_\Omega\varrho(\vv{\cdot}\nabla)\vv{\cdot}\vv\,\d x
=-\int_\Omega\varrho\vv{\cdot}{\rm div}(\vv{\otimes}\vv)\,\d x
\\[-.3em]\nonumber&\qquad
=-\int_\Omega\varrho|\vv|^2{\rm div}\,\vv+
\varrho\nabla\vv{:}(\vv{\otimes}\vv)\,\d x
\\[-.3em]&\qquad\quad
=-\int_\Omega\frac\varrho2|\vv|^2{\rm div}\,\vv\,\d x
=\int_\Omega\ff_{\rm str}{\cdot}\vv\,\d x\,,
\end{align}
where the boundary condition $\nn{\cdot}\vv=0$ and
the spatial constancy of $\varrho$ have been exploited.
In this way, the energetics \eqref{energy+} and 
\eqref{energy+++} is preserved.

Having $\varrho$ constant, the buoyancy effects must be modelled
also in a simplified way by putting
$(1{-}b(\theta))\ff$ with some increasing function $b(\cdot)$
instead of $\ff$ into \eqref{eq2} and \eqref{energy+}. This
is referred as  {\it Oberbeck-Boussinesq's approximation}. Then the
right-hand side of \eqref{eq7} is to be augmented by the adiabatic
heat source/sink $\varrho b(\theta)\ff{\cdot}\vv$. If the buoyancy
$b$ equals 0 for $\theta=0$, then $\theta\ge0$ is again guaranteed.

One of the property of the full model \eqref{eq}
is difficulty of a rigorous mathematical analysis
as far as existence of solution of the system \eqref{eq}
with the boundary conditions \eqref{BC} in
some reasonable sense. If performed in a constructive way by
some approximation, such a mathematical analysis usually
suggests also computational implementable approximation
strategies which are numerically stable and convergent. This
justifies theoretical studies also of geophysical models,
although it is mostly ignored. Unfortunately, such an analysis
of the incompressible model \eqref{eq} seems not available and
is likely very nontrivial unless some higher-gradient modifications
are adopted.

Various gradient enhancements of the plain semi-compressible model
(i.e.\ without internal variables) has been devised in \cite{Roub20QSF}
where also various impacts on dispersion of elastic (seismic) velocities
are discussed. The particular
enhancement 
by dissipative gradient terms exploits
the general ideas of {\it multipolar} (also called non-simple) media
in \cite{GreRiv64MCM} adopted for {\it fluids}
in
\cite{BeBlNe92PBMV,BeNeRa99EUFM,FriGur06TBBC,Neca94TMF,NeNoSi89GSIC,NecRuz92GSIV}. More specifically, we use 
nonlinear 2nd-grade
nonsimple fluids, also called {\it bipolar fluids},
which 
uses the dissipation potential $\zeta_1$ expanded as 
$\zeta_1(\EE(\vv))+\nu
|\nabla\EE(\vv)|^p/p$ with some coefficient $\nu>0$ and some $p>3$. Then 
the viscous stress is augmented as ${\bm S}_\text{\sc v}^{}=\zeta_1'(\EE(\vv))
-{\rm div}(\nu|\nabla\EE(\vv)|^{p-2}\nabla\EE(\vv))$
and also the dissipation rate $\xi$ now involves an additionally
heat source $\nu|\nabla\EE(\vv)|^p$.
Moreover, as \eqref{eq2+} would become a 4th-order
parabolic equation, the boundary conditions \eqref{BC1} have
to be augmented by one more higher-order condition. 

It should be mention as a general observation that
the present state of art in applied-science literature 
is that many continuum-mechanical models likely do not admit any
solutions in whatever reasonable generalized sense,
although computational simulations of certain approximate
variants are successfully launched for special data. 
Anyhow, having mere existence of solutions and
possible convergence of approximate problems 
is of theoretical interest.

Here, let us only briefly mention that, in the semi-compressible
variant, a rigorous mathematical analysis of the
system (\ref{eq}c--g) with \eqref{eq2+}
with ${\bm S}_\text{\sc v}^{}$ and $\xi$ augmented by
the multipolar terms as mentioned above
can be performed by merging and modified the results
available for an anisothermal model with damage \cite{Roub??SPTC}
with the isothermal diffusion model \cite{RouTom??CMPE}.
The essential point is to have boundedness of the velocity
gradient $\nabla\vv(t)$ in space at particular time instants $t$
and sufficiently smooth initial conditions.
The variant of a stored energy nonconvex in term of $\Ee$
like  \eqref{Vladimir} can be analytically handled
easily except that certain modification for large
$|\Ee|$ is desirable to avoid nonphysical fall of
energy to $-\infty$ if $|\Ee|\to\infty$.

\section{Conclusion}\label{sec-concl}
Let us summarize that a mechanically and thermomechanically
consistent model was devised with the goal to improve and
complete several existing models previously occurring in literature.
The main applications are presumably to tectonic and volcanic processes as
well as sources of seismic waves and their propagation in  the crust and
 the upper mantle, although it is not limitted to those. The unified
description of these processes thus allows for their natural coupling.

The Eulerian formulation, being more natural in the context of absence of
any natural reference configuration, also enables  for   an easy
enhancement of the model by interaction of some spatial fields
towards global-type models, as gravitational
(considered in Sect.\,\ref{sec-specific}) or { geo{}}magnetical applied
to paleomagnetism in rocks.
 In the latter case, interestingly the transport of the dipole
momentum (i.e.\ the magnetization) leads to a skew-symmetric
contribution to the structural stress, cf.\ \cite{Roub??TMPE}. 
Also coupling with basic models in hydrosphere (with $\chi$ representing
salinity of water in oceans) is possible to model interaction with lithosphere
during propagation of seismic waves or tsunamis caused by uplifts of
sea beds within earthquakes. 
Actually, the diffusant content $\chi$ can be vector-valued in some
applications, specifically  if several constituents are moving independently
through porous rocks.
This can be in particular water and CO$_2$, as in \cite{NorCel12GSCO},
or water and oil.  The ``monolithic'' description of 
visco-elastic  solid and fluidic areas allows for a simple modelling of
seismic waves propagation in such heterogeneous media, leading to 
reflection and refraction on the solid/fluid interfaces as
well as transformation between longitudinal (pressure) waves
and shear waves, cf.\ \cite{RouVod19MMPF} for a small-strain
variant of the model. The melting/solidification and flow/creep
modelling can be applied globally on the mantle, i.e.\
including its deeper parts (astenosphere) and lower mantle with
plumes of hot material or with falling cold slabs, not being confined
only on the crust and the subcrustal lithosphere. A more general free
energy $\psi$ may allow for phase-transition modelling in volumetric part
during recrystalization on upper/lower mantle interface. 
Also nonnegligible superheating/supercolling effects during magma solidification
(cf.\ \cite{Hamm08ESKE}) can be incorporated into the model -- see
\cite{Roub??SPTC} for one option in this direction. 

As far as magma models, there exist {\it two-phase magma dynamics
models} which consider rocks and magma as a mixture. A simple example is 
\cite{KeMaKa13MMMD} which however is fully incompressible (i.e.\ no
pressure waves can propagate) and no melting or solidification 
is, in fact, allowed and energy conservation is not ensured.
For a compressible or a multi-component variant see
\cite{DanHei16CMMD,Gery19INGM,YarPod15DPVM} or 
\cite{KelSuc19CMMP}, respectively. Other models by \cite{BeRiSc01TPMC} or 
\cite{SrRiBe07SMCD} allow for 
melting/solidification and 
consider energy conservation in the continuous model
but are again incompressible. Balancing momenta of each
constituents separately is a rational approach
usually credited by \cite{True68BTM}, working
satisfactorily rather for two-component mixtures only,
cf.\ \cite{RajTao95MM}, in contrast to a 
phenomenological Nobel-prize awarded approach by \cite{Prig62ITIP}.
Cf.\ \cite{Samo07ATMM} for a comparison of these mixture approaches.
In comparison, our model is
``monolithic'', reflecting the fact that rocks and magma
are actually the same material which is only in different mechanical
state depending mainly on temperature and pressure, anyhow allowing,
in addition, for a possible phenomenological diffusion of some constituents
(e.g.\ water, hydrated minerals, or oil) and in addition for a fracture in the
solid regions (which may lead to volcanic earthquakes) and for a pressure wave
propagation.
In fact, the mentioned diffusion which resulted by the Biot model
is a flow due to Fick or Darcy law (cf.\ Sect.\,3.6 in
\cite{KruRou19MMCM}) and represents a simplest variant of
the hierarchy of several porous-media models, cf.\ \cite{Raja07HAMF}.

An alternative to the semi-compressible simplification from
Section~\ref{sec-semi}, where the inertia is kept in the model but
mass density variations are neglected both in time and space (i.e.\ 
$\varrho$ is constant), can be just opposite: inertia to be
neglected by considering zero acceleration $\DT\vv=0$
but mass density $\varrho$ to allow to vary according to some state
equation, being a function of pressure and temperature, cf.\
\cite{Gery19INGM}.
Of course, such {\it quasistatic} (or sometimes called {\it quasi-dynamical})
 {\it models} suppress not only seismic pressure wave
propagation (like in incompressible variants) but also shear waves. On the other
hand, they can still cover many medium- and long-time scale phenomena
in a computationally efficient way, cf.\ e.g.\
\cite{BeuPod10VMCL,BeRiSc01TPMC,DanHei16CMMD,GerYue07RCMM,JacCac20MMBD,KeMaKa13MMMD,KrHeBa12HAMC,PCTC17SMFV,ThKaPo15LSRB}.

In the quasistatic variant, the {\it buoyancy} model
of the {\it Oberbeck-Boussinesq type} as mentioned in Section~\ref{sec-semi}
has been used e.g.\ in \cite{DanHei16CMMD,JacCac20MMBD,KrHeBa12HAMC} in a
thermodynamically consistent variant including adiabatic effects or
also in e.g.\ in \cite{BeuPod10VMCL,GerYue07RCMM,PCTC17SMFV,ThKaPo15LSRB}
with adiabatic effects neglected. These quasistatic buoyancy models, in fact,
consider the mass density temperature dependent.

Eventually, let us conceptually discuss some implementation issues.
It has been articulated in \cite{DMGT11DEFS} that
``computational modeling in geology requires numerical methods
that are robust, reliable and accurate''. The robustness and reliability
should be supported by rigorous mathematical proofs of
existence of solutions to the particular models and of
data-depending bounds of solutions obtained by specific
approximation methods, leading (under some assumptions on
the data) to guaranteed numerical stability and convergence of
particular discretization methods.
Even more demanding numerical analysis might yield estimates of 
rates of discretization errors, although this is rarely
available. For very practical reasons, such theoretical justification is
largely ignored in computational (geo)physics as well as in engineering,
being substituted by particular computational simulations and
comparison with available experiments, which however, strictly speaking,
cannot be considered
as a really rigorous proof of robustness and reliability. Many models
used in geophysics even have no guaranty for mere existence
of solutions in any reasonable sense.

At this occasion, it should be mention that, in general,  the
consistent  physically motivated 
energetics  like this one presented here in
Sect.\,\ref{sec-termodynam} opens possibilities
 
for designing numerical{ly } stable and convergent approximation
schemes.  This can be a nontrivial task in particular models, however.
Here, more specifically, the semicompressible variant from
Section~\ref{sec-semi} bears fully implicit time discretization (i.e.\ the
backward Euler method with possible
semi-implicitly handled coefficients in dissipative terms),
cf.\ \cite{Roub??SPTC} and, for fine numerical aspects of fully implicit
time discretization of problems with nonconvex energies, also \cite{Roub20CTDD}.
Notably, the objective time derivatives in (\ref{eq}c--g) do not allow for
staggered time discretization, which in the non-convective variant
would be efficient and even energy conserving, cf.\ \cite{Roub17ECTD}.

Of course, the fully implicit time and space discretization leads to
nonlinear algebraic systems at each time level to be solved only iteratively,
which might be algorithmically rather demanding in situation
when fast evolving processes (in particular during rupture and
earthquakes) would occur. Beside, some adaptive refinements
both of time discretization and space discretization would be
desired (e.g.\ fine time discretization is needed only during
ongoing ruptures and propagation of seismic waves  
and, in most situations, damage is localized only around existing
faults). The adaptively varying time step is well facilitated by one-step
discretization and, except explicit time discretization methods,
the stability of the resulted scheme is not corrupted by large time steps.
Such algorithmic and implementation issues would be surely very
demanding especially in full model and 
they have been out of scope of this theoretical article, as well as 
launching some computational simulations even for some partial, simplified
variant of this model.

Algorithmically, truly efficient calculations of seismic waves
need an explicit time discretisation at least of the elastodynamical
part, cf.\ \cite{RouTso21SEIT}. Of course, these time discretisations are to
be combined with space discretization by some standard methods like
finite or spectral elements, although the objective time derivatives
again makes the numerical analysis very technical because they do not
stay functions in the respective finite-dimensional spaces. 
The calculations
from Section~\ref{sec-termodynam} leading to the desired energetics thus
could not be executed directly for the spatially discretized problem and
would have to be done successively, likely with other regularizing gradient
terms. Even more, the explicit time discretizations work successfully for
hyperbolic-type problems, which would here require to suppress
the dissipative stress ${\bm S}_\text{\sc v}^{}$ but this would
corrupt the theoretical arguments behind existence of solutions
of the model mentioned in Section~\ref{sec-semi}. The fully compressible model
with the continuity equation \eqref{eq} would be even more
delicate and combination with discretization methods for mere compressible
fluids as \cite{DolFei15DGMA,FeKaPo16MTCV}
should be elaborated. Rather, it is expected that only the mentioned quasistatic
variant of the model \eqref{eq} bears really efficient
computational implementation.

\bigskip

\baselineskip=10pt
\centerline{\it Acknowledgments.} 
{\small

\medskip

\noindent
The author is deeply thankful to Dr.\ Vladimir Lyakhovsky 
for inspiring discussions in 2014 during author's visit of the
Geological Survey of Israel, and to two referees for critical reading
of the original version and many valuable suggestions.
Also the support from the M\v SMT \v CR (Ministry of Education of the
Czech Republic) project CZ.02.1.01/0.0/0.0/15-003/0000493
and the institutional support RVO:61388998 (\v CR) are acknowledged.
}





\end{sloppypar}
\end{document}